**Probing the Mind Behind the (Literal and Figurative) Lightbulb**

Liane Gabora

University of British Columbia

A requested commentary on "Thomas Edison's creative career: The Multilayered trajectory of trials, errors, failures, and triumphs" by Dean Simonton

x

Abstract

After doing away with the evolutionary scaffold for BVSR, what remains is a notion of "blindness" that does not distinguish BVSR from other theories of creativity, and an assumption that creativity can be understood by treating ideas as discrete, countable entities, as opposed to different external manifestations of a singular gradually solidifying internal conception. Uprooted from Darwinian theory, BVSR lacks a scientific framework that can be called upon to generate hypotheses and test them. In lieu of such a framework, hypotheses appear to be generated on the basis of previous data—they are not theory-driven. Simonton (2014) does not explain how the hypothesis that creativity is enhanced by engagement in a "network of enterprises" is derived from BVSR; this hypothesis is more compatible with competing conceptions of creativity. The notion that creativity involves backtracking conflates evidence for backtracking with respect to the external output with evidence for backtracking of the conception of the invention. The first does not imply the second; a creator can set aside a creative output but cannot go back to the conception of the task he/she had prior to generating that output. The notion that creativity entails superfluity (i.e., many ideas have "zero usefulness") is misguided; usefulness is context-dependent, moreover, the usefulness of an idea may reside in its being a critical stepping-stone to a subsequent idea.

**Probing the Mind Behind the (Literal and Figurative) Lightbulb**

Simonton's paper appears to serve two purposes: to provide an analysis of the patent output of Thomas Edison, and to interpret this analysis as support for the BVSR theory of creativity. It has considerable merit as a carefully executed and indeed fascinating account of the lifetime output of a famous creator. As support for a theory of how the creative process works, however, it is not clear what it contributes. This commentary addresses the soundness of the theory itself and the interpretation of the Edison findings as support for the theory.

**BVSR Today**

BVSR was conceived of as a Darwinian theory (Campbell, 1960), and the strength of BVSR was that, if true, it would place creative processes in a theoretical framework that accounts for the evolution of life through natural selection, and thus provide a unified account of how new forms, both biological and creative, arise and evolve over time. If successful, this would have been a considerable accomplishment. However, after meeting with criticism, Simonton has backed away from Darwinism as a theoretical framework. He recently claimed that his theory has been "radically reformulated" to show that "BVSR's explanatory value does not depend on any specious association with Darwin's theory of evolution" (Simonton, 2012). The current paper continues this trend of steering clear of a Darwinian framework while continuing to employ the phrase BVSR, and in particular the concept of "blindness".

What is meant by "blindness" has shifted over the years. Simonton invokes a definition from a paper published last year in which the term is used to indicate, "only that the person be willing and able to generate low-probability (i.e., high originality) ideas without solid prior knowledge of whether any given idea will actually prove useful or not" (Simonton, 2013). This definition strikes me as tautological. Obviously creative ideas are generated by people who are willing and able to generate original ideas, since originality is a component of creativity, and obviously, since the space of creative possibilities is open-ended, one never knows in advance the possible uses or applications of an idea. If this is now all that is meant by "blindness" I know of no theory of creativity that is inconsistent with it. Thus "blindness" is no more supportive of BVSR than of any other theory.

Here is a summary, as best as I can make out, of what, in addition to the notion of blindness, the BVSR theory now comprises:

- A commitment to the notion that ideas can be treated as distinct, countable entities (as opposed to interrelated manifestations of a singular solidifying internal conception)
- A claim to compatibility with the notion of creativity as a "network of enterprises"
- Tell-tale characteristics
    - Superfluity
    - Backtracking

I will address each of these in turn, following a brief discussion of the working hypothesis underlying the paper.

**The Working Hypothesis**

On p. 6 the author writes (with text I refer back to bolded by me):

> "The working hypothesis is that the sum total of his 1093 patented inventions will prove compatible with the patterns that have just been suggested. **Good ideas will be mixed up with bad ideas**, and **some trains of thought will lead nowhere creative**. Edison was able to create so many first-rate inventions largely because he was able to conceive so **many ideas that had no merit whatsoever**."

**Good Ideas will be Mixed up with Bad Ideas**. It is unfortunate that Simonton does not explain what he means by "good ideas will be mixed up with bad ideas". Formal treatments of concept combination are showing that there are many fascinating ways in which concepts or ideas can be (to use his term) "mixed up" (e.g. Aerts, Gabora, & Sozzo, 2013; Thagard & Stewart, 2011), and the implications of such models for creativity are being explored (e.g., Veloz et al, 2011). However, the further one reads the more one gets the impression that he is not referring to the all-important question of how concepts merge to form new ideas; by "mixed up" he means "interspersed", i.e., a good idea may be followed by a bad idea which may be followed by a good idea. The terms "good" and "bad" imply that there exists some sort of objective evaluation function over the space of ideas, but in fact an idea has merit only *with respect to* a particular context, i.e., a particular need or aesthetic sensibility. (For readers who are familiar with attempts by myself and others to formalize these notions, this amounts to saying that an idea does not have value until it "collapses with respect to a particular measurement"). An idea that is of no merit with respect to one context may be of merit with respect to another. Indeed, ideas that are most ahead of their time are of no value during the creator's lifetime because their implications take so long to be realized. Ideas that do not work on our planet (with its particular atmosphere, gravitational force, and so forth) may one day work perfectly on another planet. In short, an evaluation of "good" or "bad" is as much a reflection of the perspective it is evaluated from as of the ideas itself. What Edison may have been doing was optimizing with respect to one constraint by considering the idea from one perspective, and then optimizing with respect to another constraint by considering it from another perspective, and through this kind of tinkering he eventually reached a pareto optimal solution (such that he could no longer improve with respect to one constraint without getting worse with respect to another). There may have been a bit of trial and error involved but I suspect the tinkering was less a matter of crossing off "bad" ideas than of incrementally gaining a richer conception of the task and how to go about it.

**Some Trains of Thought will Lead Nowhere Creative.** It is clear from the phrasing "nowhere creative" that Simonton defines and measures creativity with respect to the change it exerts in the external world. An alternative is it be defined and measured in terms of the change it brings about in the minds of the creator and appreciators of the creative work; thus a train of thought does not have to *lead to* somewhere creative to *be* creative (Gabora, O'Connor, & Ranjan, 2012; Ranjan, 2014). If a particular train of thought transformed Edison's conception of the problem then it *was* creative, whether or not it resulted in a creative outcome. A theory of creativity that focuses on external results cannot explain common attitudes toward creative artifacts, such as that while an original masterpiece is viewed as creative, a reproduction or imitation of it is not. A theory that focuses on internal change can make sense of this; only the original masterpiece provided humanity with a newfound roadmap to understanding or expressing something.

**Many Ideas had no Merit Whatsoever**. Since the space of creative outputs is open-ended, it is not possible to say that any of Edison's ideas "had no merit whatsoever". An idea that is without merit when expressed in the world may nevertheless be a crucial step forward toward a subsequent conception that *does* manifest as a creative output. The merit of an idea can derive not just from what it directly generates in the external world but from the shift it brings about in the creator's understanding.

### The Notion that Ideas can be Treated as Distinct, Countable Entities

There is mounting evidence that one cannot rely on finished outputs to study creative processes; this is in part the motivation for studies in which individuals are stopped midway through their creative process (Bowers et al, 1995; Gabora & Saab, 2011; Carbert, Gabora, Schwartz, & Ranjan, 2014). This is particularly true if the adaptive landscape associated with the task is peaked (multiple routes to the same outcome). Enter Simonton who brazenly makes claims about creative processes not just on the basis of careful examination of creative products, but on the basis of simply adding up how many ideas were patented in a given year.

Simonton's historiometric analyses of creative individuals are unparalleled, and (other than his Darwinian / BVSR arguments) I believe one could argue he as contributed as much or more to the psychology of creativity as anyone. However, his mathematical treatment of creativity, both here and elsewhere, is limited, indeed arguably trivial, for it in no way takes into account the actual structure of an idea nor how previous or subsequent conceptions are related to it. Simonton treats ideas as discrete, distinct, countable entities, and his writing is permeated with phrasing that indicates this is how he really conceives of them, e.g., the phrase "sheer number of" in "The whole episode illustrates the sheer number of *blind* ideas he was willing to generate and test." Let us examine the implications of starting off with the assumption that each idea can be treated as a separate entity without regard to its interrelationships to other ideas. Say one has to go through 100 steps to get to an idea, which we call *idea 1*. Let us say that *idea 1* can be applied, in one step each, to two different fields of knowledge resulting in two different outcomes: *c1* and *c2*. The step of applying *idea 1* in one of these two ways is step 101. Say further that one has to go through 101 completely different steps to get to a third creative outcome, *c3*. The approach taken here treats *c1, c2,* and *c3* as equally similar; it ignores that while *c3* is completely different, *c1* and *c2* share a common core. Without going into details, this interrelatedness is why in the honing theory of creativity trials are characterized as different *actualizations* of an unfolding, underlying conception that exists, initially, in a state of potentiality, and that emerges in an ecology of related ideas.

Simonton claims that because creativity involves venturing into the unknown it "requires the introduction of trial and error procedures", by which he means, "many ideas must be generated, but few will be chosen". However, trial and error does not just provide the creator with information that a particular idea doesn't work, it better acquaints the creator with the lay of the land in such a way as to provide hunches that guide subsequent efforts. Given that, as a consequence of the distributed, content-addressable nature of associative memory, "like attracts like" (even if the "likeness relation" has never been noticed before) it would be a huge waste if creativity were merely a matter of ticking off each trial that doesn't pan out one by one and moving on to the next (Gabora, 2000, 2010). It is much more consistent with the architecture of memory that the creator's conception of the task incorporates all sorts of potentially related knowledge and ideas into a single ill-defined entity that becomes better defined through the process of considering it from different perspectives.

### "Network of Enterprises" is More Consistent with Alternative Theories

In order that a hypothesis support a theory of creativity it is necessary to show why the theory gives rise to the hypothesis. With respect to the hypothesis that creativity is enhanced by engagement in a network of enterprises", for which there was already considerable empirical support prior to this paper, not only is it not explained how it is derived from the theory, it is actually inconsistent with the analysis carried out in this paper. As we have seen, the analysis is based on the assumption that one can make claims about creativity while treating ideas as distinct, countable entities, ignoring interrelationships amongst them, even if they are different manifestations of a common underlying core. The network of enterprises notion is actually much more consistent with competing theories such as honing theory according to which it is necessary to take into account the inner structure of ideas, their relationships, and their relative positions with respect to a larger ecological structure of ideas.

On a related note, Simonton writes of a "residual uniqueness" that defined Edison and made him a creative genius. However, it is not clear how a theory that emphasizes the mere generation of ideas and treats them as discrete, separate entities can hope to achieve this. On the other hand, residual uniqueness, or recognizable personal style, arises naturally when creative outputs are viewed as manifestations of a unique, self-organizing worldview (Gabora, O'Connor, & Ranjan, 2012; Ranjan, 2014).

## "Tell-tale Characteristics" are Not Diagnostic of BVSR

The paper claims there are two "tell-tale" characteristics of BVSR: superfluity and backtracking. We now examine this claim.

**Superfluity**

The author defines the first "tell-tale" characteristic as follows:

> *Superfluity* means that the creator generates more ideas than are strictly necessary in the sense that many if not most of those generated have zero usefulness (or at least lower utilities than the ideas that are eventually found to have the highest utility). Logically, if we assume that creativity is rational, then the only reason why a relatively useless idea would ever be generated and tested is because its comparative or absolute utility is not already known in advance. That prior ignorance interjects the requisite *blindness* behind BVSR. Only through trial and error can the creator determine the idea with the maximum utility.

If this passage were true then artists would not generate sketches and engineers would not build toy models, for they know in advance that these will not be the final, useful outputs. Simonton's phrasing here again reveals his reliance on external final outcomes in making claims about internal processes. Particularly in light of work on the intuitive antecedents of insight (e.g., Bowers et al, 1995) it seems likely that creators generate and test ideas that they know will not be the final product because they intuitively sense that these ideas are vital stepping stones toward the final product. Simonton also implies here that a creator can transit from a state of not knowing the utility of an idea to a state of knowing it, but as discussed above, the utility of an idea can never be objectively known; it can only be determined with respect to a particular context.

Simonton then draws upon a quote from Edison to support the notion that creativity involves superfluity:

> Edison exclaimed, "Why, man I have got a lot of results. I have found several thousand things that won't work" (Whitehorne, 1920, p. 258). This comment illustrates how the inventor knew that failures add increments to the accumulation of knowledge.

Clearly learning that something failed adds to ones' knowledge base, but I do not think this is what Edison was driving at with this comment. He was expressing that his failures had whittled away at the potentiality of an initially vague conception, rendering it more concrete. Each "trial" did not just tell Edison that a particular idea was or was not effective, it re-shaped his understanding of the problem and issues he faced in solving it. New information affects creativity not by merely accumulating but by transforming the state space; indeed it can render whole bodies of previous knowledge obsolete. The bottom line here is that cherry-picked offhand comments that are interpreted as telltale signs of BVSR are not diagnostic; they can equally be interpreted as support for competing views of how creativity works.

**Backtracking of Outputs Does Not Imply Backtracking in Conception of Task**

Simonton conflates evidence for backtracking with respect to the external output with evidence for backtracking of the conception of the invention. The first does not imply the second; a creator can set aside a particular output but cannot go back to the conception of the task he/she had prior to generating that output. I can think of no theory of creativity that is incompatible with backtracking with respect to the *product*. Therefore findings of backtracking with respect to the product are not diagnostic of BVSR.

In a reply to a commentary Simonton wrote on a paper of mine (Gabora, 2013), I suggested that his previous findings of backtracking, "nonmonotonicity", and so forth could be indicative of self-organized criticality as predicted by a competing theory, honing theory. He could have used the Edison data set to test that hypothesis but he would have had to take the further steps of getting objective judges to rate *how* creative each patented idea was, and plot the frequency of ideas with each possible rating on a log-log plot, but this he did not do. This is unfortunate because such a test (though still suffering from the limitation of relying solely on finished outputs) could have been of diagnostic value.

### Other Minor Issues

In both the abstract (p. 2) and the discussion section (p. 26) the author uses the phrase "some form of BVSR". If there are multiple forms of BVSR he is encouraged to either explain these alternative forms or provide a reference to where they are explained. Otherwise the phrase comes across as a catch-all, i.e., it seems to be saying, "I am going to remain vague about my theory, implying that there are many possible forms of it without explaining what they are so that it will be possible to interpret any forthcoming findings as consistent with some form of my theory".

Simonton writes, "In brief, the creator must take risks, even risking utter failure. Expertise alone is insufficient to guarantee creative success." Does anyone really believe that a creator need not take risks, or that expertise guarantees success? If he has reason to believe that these statements are at all controversial he should cite the relevant literature.

A final small point: Simonton makes a big deal out of Edison's breaking of the supposed ten-year rule since he started to procure patents after only seven years apprenticeship, not ten. However, many creative individuals (perhaps the majority) do not manage to make a living doing what they eventually become know for, and can only work on creative projects in their spare hours, whereas Edison was able to work full-time on projects directly related to his creative interests. Thus although Edison's successes came earlier, the total number of hours he put in before procuring patents may be about average. Note also that the bulk of his successful creative output did occur at the ten-year mark.

## Conclusion

Although I found the historical, biographical aspect of this paper fascinating, scientifically I found it unsatisfying. The rich complexity of Edison's creative process has been distilled to a set of 1s and 0s: a full-fledged idea either got patented or not. In fact, ideas that did not get patented are only mentioned in passing; they are not included in the analysis. Clearly there are severe limitations to what such an analysis can yield about creativity. The paper is additionally plagued by other problems. Uprooted from Darwinian theory, BVSR lacks a scientific framework that can be called upon to generate hypotheses and test them. In lieu of such a framework, hypotheses appear to be generated on the basis of previous data—they are not theory-driven—and data supporting the hypotheses is then put forward as evidence of BVSR. It is perhaps unfortunate that in previous invited commentaries on papers by this author and other papers featuring discussion of BVSR (Gabora, 2005, 2007, 1010, 2011) I focused on its failure to satisfy the requirements of a Darwinian theory, since that appeared to be its most glaring problem. The Darwinism issue stems from a deeper issue: the gulf that separates 'many are generated and few selected" modes of change from "actualization of potential" modes of change, which, although its implicates permeate psychology, I became aware of only through collaboration with mathematical physicists. If, instead of going to contorted lengths to salvage shreds of the sinking BVSR ship, Simonton had investigated this (as I suggested to him 15 years ago), I believe it would soon have become clear to him which mode is appropriate for a theory of creativity.

All in all I was left with the impression that his belief that creative success is associated with "the sheer number of *blind* ideas [one is] willing to generate and test" has led him to generate a large number of papers thereby increasing his odds of getting it right. This would be an effective strategy if playing this kind of numbers game really captures what creativity is about, but I don't think it does; I think creativity is about whittling away at potentiality and transforming a space of possibilities.

In short, this paper sheds more light on the story of the external, physical lightbulb than it does on the internal process that is sometimes signified by a lightbulb. Nevertheless, the author's research record is impressive and I hope that his next paper points the way to a more penetrating conception of how creativity works.

**Acknowledgements**

MIND BEHIND THE LIGHTBULB                                              14This research was conducted with the assistance of grants from the National Science and Engineering Research Council of Canada, and the Fund for Scientific Research of Flanders, Belgium.**References**

Aerts, D., Gabora, L., & Sozzo, S. (2013). Concepts and their dynamics: A quantum-theoretic modeling of human thought. *Topics in Cognitive Science*, *5*(4), 737-772.

Bowers, K. S., P. Farvolden, & L. Mermigis. (1995). Intuitive Antecedents of Insight. In S. M. Smith, T. B. Ward, & R. A. Finke (Eds.). *The creative cognition approach* (pp. 27-52). Cambridge MA: MIT Press.

Campbell, D. T. (1960). Blind variation and selective retention in creative thought as in other knowledge processes. *Psychological Review, 67,* 380−400.

Carbert, N., Gabora, L., Schwartz, J., & Ranjan, A. (2014). States of cognitive potentiality in art-making. *Proceedings of the AIEA Congress on Empirical Aesthetics*. Held August 22-24, New York. Rome, Italy: International Association of Empirical Aesthetics.

Gabora, L. (2000). Toward a theory of creative inklings. In (R. Ascott, Ed.) *Art, technology, and consciousness* (pp. 159–164). Bristol UK: Intellect Press.

Gabora, L. (2005). Creative thought as a non-Darwinian evolutionary process. *Journal of Creative Behavior*, *39*(4), 65–87.

Gabora, L. (2007). Why the creative process is not Darwinian. Commentary on D. K. Simonton 'The creative process in Picasso's Guernica sketches: Monotonic improvements versus nonmonotonic variants.' *Creativity Research Journal*, *19*(4), 361–365.

Gabora, L. (2010). Why Blind-Variation-Selective-Attention is inappropriate as an explanatory framework for creativity. *Physics of Life Reviews, 7*(2), 190-194.

MIND BEHIND THE LIGHTBULB                                                                                                    15

Gabora, L. (2011). An analysis of the Blind Variation and Selective Retention (BVSR) theory of creativity. *Creativity Research Journal, 23*(2), 155–165.

Gabora, L., O'Connor, B., & Ranjan, A. (2012). The recognizability of individual creative styles within and across domains. *Psychology of Aesthetics, Creativity, and the Arts*, *6*(4), 351-360.

Gabora, L. & Saab, A. (2011). Creative interference and states of potentiality in analogy problem solving. *Proceedings of the Annual Meeting of the Cognitive Science Society* (pp. 3506-3511). Held July 20-23, Boston MA. Austin TX: Cognitive Science Society.

Ranjan, A. (2014). *Understanding the creative process: Personal signatures and cross-domain interpretations of ideas*. Ph.D. Thesis, University of British Columbia, Canada.

Simonton, D. K. (2011). Creativity and discovery as blind variation: Campbell's (1960) BVSR model after the half-century mark. *Review of General Psychology*, *15*, 158-174.

Simonton, D.K. (2012). Creativity, problem solving, and solution set sightedness: Radically reformulating BVSR, *Creativity Research Journal, 46*, 48–65.

Simonton, D. K. (2013). Creative thought as blind variation and selective retention: Why sightedness is inversely related to creativity. *Journal of Theoretical and Philosophical Psychology*, *33*, 253-266.

Simonton, D. K. (2014). Thomas Edison's creative career: The multilayered trajectory of trials, errors, failures, and triumphs. *Psychology of Aesthetics, Creativity, and the Arts*. Advance Online Publication. doi: 10.1037/a0037722

Thagard, P., & Stewart, T. C. (2011). The AHA! experience: Creativity through emergent binding in neural networks. *Cognitive Science, 35,* 1–33.

Veloz, T., Gabora, L., Eyjolfson, M., & Aerts, D. (2011). A model of the shifting relationship between concepts and contexts in different modes of thought. *Lecture Notes in Computer*

*Science 7052: Proceedings of the Fifth International Symposium on Quantum Interaction*. June 27-29, Aberdeen, UK. Berlin: Springer.

Whitehorne, E. E. (1920). Edison–Wizard in business. *Electrical Merchandizing, 28,* 226–230. 258.